\documentclass[
 reprint,
 groupedaddress,
 showpacs,
 superscriptaddress, 
 amsmath,amssymb,
 aps,
 prl,
 nofootinbib,
 twocolumn,
 pdflatex
]{revtex4-1}

\usepackage{setspace} 
\usepackage{amssymb}
\usepackage{amsmath}
\usepackage{physics}
\usepackage{empheq}
\usepackage{amsthm}
\usepackage{mathtools}
\usepackage{hyperref}
\usepackage[normalem]{ulem}  
\usepackage[whole]{bxcjkjatype}
\usepackage{amssymb}
\usepackage[pdftex]{graphicx}
\usepackage[short]{turnthepage}
\usepackage{comment}
\usepackage{footmisc}

\hypersetup{
    colorlinks=true,
    citecolor=magenta,
    linkcolor=magenta,
    urlcolor=blue,
}

\mathtoolsset{showonlyrefs=true}

\theoremstyle{definition}

\newcommand{\uu}{\vb*{u}}

\newcommand{\rr}{\vb*{r}}

\newcommand{\pp}{\vb*{p}}

\newcommand{\beb}{\begin{itembox}}
\newcommand{\enb}{\end{itembox}}

\newcommand{\e}{\varepsilon}
\newcommand{\E}{\vb*{E}}
\newcommand{\B}{\vb*{B}}

\allowdisplaybreaks

\begin{document}

\title{Plasmonic quantum nonlinear Hall effect in noncentrosymmetric 2D materials}

\author{Riki Toshio}
\email{toshio.riki.63c@st.kyoto-u.ac.jp}
\affiliation{%
 Department of Physics, Kyoto University, Kyoto 606-8502, Japan
}%

\author{Norio Kawakami}
\affiliation{%
 Department of Physics, Kyoto University, Kyoto 606-8502, Japan
}%

\begin{abstract}
We investigate an interplay between quantum geometrical effects and surface plasmons through surface plasmonic structures, based on an electron hydrodynamic theory. First we demonstrate that the quantum nonlinear Hall effect can be dramatically enhanced over a very broad range of frequency by utilizing plasmonic resonances and near-field effects of grating gates. Under the resonant condition, the enhancement becomes several orders of magnitude larger than the case without the nanostructures, while the peaks of high-harmonic plasmons expand broadly and emerge under the off-resonant condition, leading to a remarkably broad spectrum. 
Furthermore, we clarify a universal relation between the photocurrent induced by the Berry curvature dipole and the optical absorption, which is essential for computational material design of long-wavelength photodetectors.
Next we discuss a novel mechanism of geometrical photocurrent, which originates from an anomalous force induced by oscillating magnetic fields and is described by the dipole moment of orbital magnetic moments of Bloch electrons in the momentum space. Our theory is relevant to 2D quantum materials such as layered WTe${}_2$ and twisted bilayer graphene, thereby providing a promising route toward a novel type of highly sensitive, broadband terahertz photodetectors.

\end{abstract}

\date{\today}

\maketitle


\textit{Introduction.---}
Quantum geometry plays a crucial role in the linear/nonlinear optical responses of bulk crystals, as exemplified by natural optical activity~\cite{Malashevich2010,Orenstein2013,Zhong2015,Ma2015,Hosur2015,Zhong2016,Flicker2018,Wang2020}, bulk photovoltaic effect~\cite{Aversa1995,Sipe2000,Nagaosa2017,Ahn2020,Watanabe2021}, and geometric photon drag~\cite{Shi2021}.
These phenomena provide us with not only a deep insight into the band structure of crystals, but also a variety of functional optical devices, such as solar cells~\cite{Tan2016,Cook2017} and infrared/terahertz photodetectors~\cite{Liu2020,Isobe2020,Zhang2021}. 
For example, recently, the quantum nonlinear Hall (QNLH) effect~\cite{Sodemann2015} ---an intrinsic low-frequency photocurrent driven by the Berry curvature dipole--- is attracting much interest as a promising candidate for a broadband long-wavelength photodetector at room temperature~\cite{Zhang2021,Du2021}.

Plasmonic nanostructures also provide us with another type of efficient and electrically-tunable optical devices~\cite{Atwater2010,Yao2014,Low2014,Hentschel2017,Yang2022}. Such a plasmonic nanodevice achieves its remarkable performance by utilizing
the nonlocality and the plasmonic enhancement triggered by the nanostructures.
In particular, surface plasmons inherent in two-dimensional (2D) layered systems, such as graphene, are known to have remarkably long lifetimes and electrically-tunable dispersions in the terahertz or mid-infrared region~\cite{Ju2011,Basov2016,Li2017}. These properties are ideal for plasmonic devices, and thus a lot of papers have been devoted to investigating the applications such as tunable terahertz photodetectors~\cite{Koppens2014,Bandurin2018_2,Ryzhii2020,Y_Zhang2021} and broadband absorbers~\cite{Ke2015,Chaudhuri2018,Huang2020}.

Electron hydrodynamics, which is quickly growing into a
mature field of condensed matter physics~\cite{Narozhny2017,Lucas2018_review,Narozhny2019,Polini2019,Narozhny2022}, gives us a powerful tool to describe electronic collective modes~\cite{Principi2016,Gorbar2017,Zhang2018,Zhang2018_2,Svintsov2018,Sukhachov2018,Gorbar2018,Narozhny2021,Kapralov2022} and nonlocality of optical responses~\cite{Raza2015,Mendoza2013,Forcella2014,Yan2015,Sun2018_2,Sun2018,Alekseev2018,Alekseev2019,Ryzhii2020,Potashin2020,Shabbir2020,Alekseev2021,Man2021,Y_Zhang2021,Luca2021,Pusep2022,Valentinis2022},
Remarkable examples related with optical applications include the theory of the plasmonic instability~\cite{Dyakonov1993,Dyakonov1996,Mikhailov1998,Vicarelli2012,Tomadin2013,Davide2014,Torre2015,Mendl2018,Mendl2021,Cosme2021,Crabb2021} and the ratchet effect~\cite{Popov2011,Olbrich2011,Ivchenko2011,Povov2013,Rozhansky2015,Olbrich2016,Rupper2018,Monch2022,Monch2022_2}, both of which harness the plasma modes to realize highly efficient photovoltaic conversions. Interestingly, for the latter, the hydrodynamic signature has been observed very recently in bilayer graphene with an asymmetric dual-grating gate potential~\cite{Monch2022,Monch2022_2}.
As a more recent development, symmetry of crystals and quantum geometry give a new twist to the concept of electron hydrodynamics~\cite{Cook2019,Varnavides2020,Rao2020,Robredo2021,Rao2021, Gorbar2018, Link2020,Varnavides2020_2,Toshio2020,Funaki2021,Funaki2021_2,Tatara2021,Tavakol2021,Hasdeo2021,Sano2021,Wang2021,Varnavides2022,Varnavides2022_2,Huang2022,Friedman2022,Valentinis2022,Sano2022}.
Indeed, in the last few years, a number of papers have addressed rich and novel hydrodynamic phenomena, represented by anisotropic viscosity effects~\cite{Cook2019,Varnavides2020,Rao2020,Robredo2021,Rao2021}. Especially for noncentrosymmetric systems, it has been revealed that the quantum geometry causes anomalous driving forces over electon fluids, leading to unique hydrodynamic phenomena such as asymmetric Poiseuille flows~\cite{Toshio2020} and nonreciprocal surface plasmons~\cite{Sano2021}. These frameworks enable us to investigate the interplay between quantum geometry and surface plasmons in novel materials, such as topological or van der Waals (vdW) materials~\cite{Hasan2010,Stauber2014,Basov2016,Yan2017}. These issues have not been executed so far, except for several limited problems~\cite{Pellegrino2015,Principi2016,Song2016,Hofmann2016,Kumar2016,Gorbar2017,Zhang2018,Zhang2018_2,Lu2021,Sano2021}.

In this Letter, based on an electron hydrodynamics, we develop a generic theory of geometrical photocurrent in noncentrosymmetric 2D layered systems with periodic grating gates.
First we demonstrate that the QNLH effect is enhanced dramatically by 
plasmonic resonances and near-field effects of grating gates, which is dubbed the {\it plasmonic QNLH effect}.
It features multiple sharp peaks near the plasma frequencies, and could be enhanced by several orders of magnitude over a very broad range of frequency. 
Furthermore, assuming more generic situations, we uncover a universal relation between the photocurrent induced by the Berry curvature dipole and the optical absorption, which is essential for computational material design of long-wavelength photodetectors.
Finally we discuss another type of novel geometrical photocurrent, the {\it magnetically-driven plasmonic photogalvanic effect}.
This is a spatially dispersive contribution to the total photocurrent
and originates from the anomalous driving force on electron fluids, which is described by the dipole moment of orbital magnetic moments in the momentum space. 

\begin{figure}
    \centering
    \includegraphics[width=8cm]{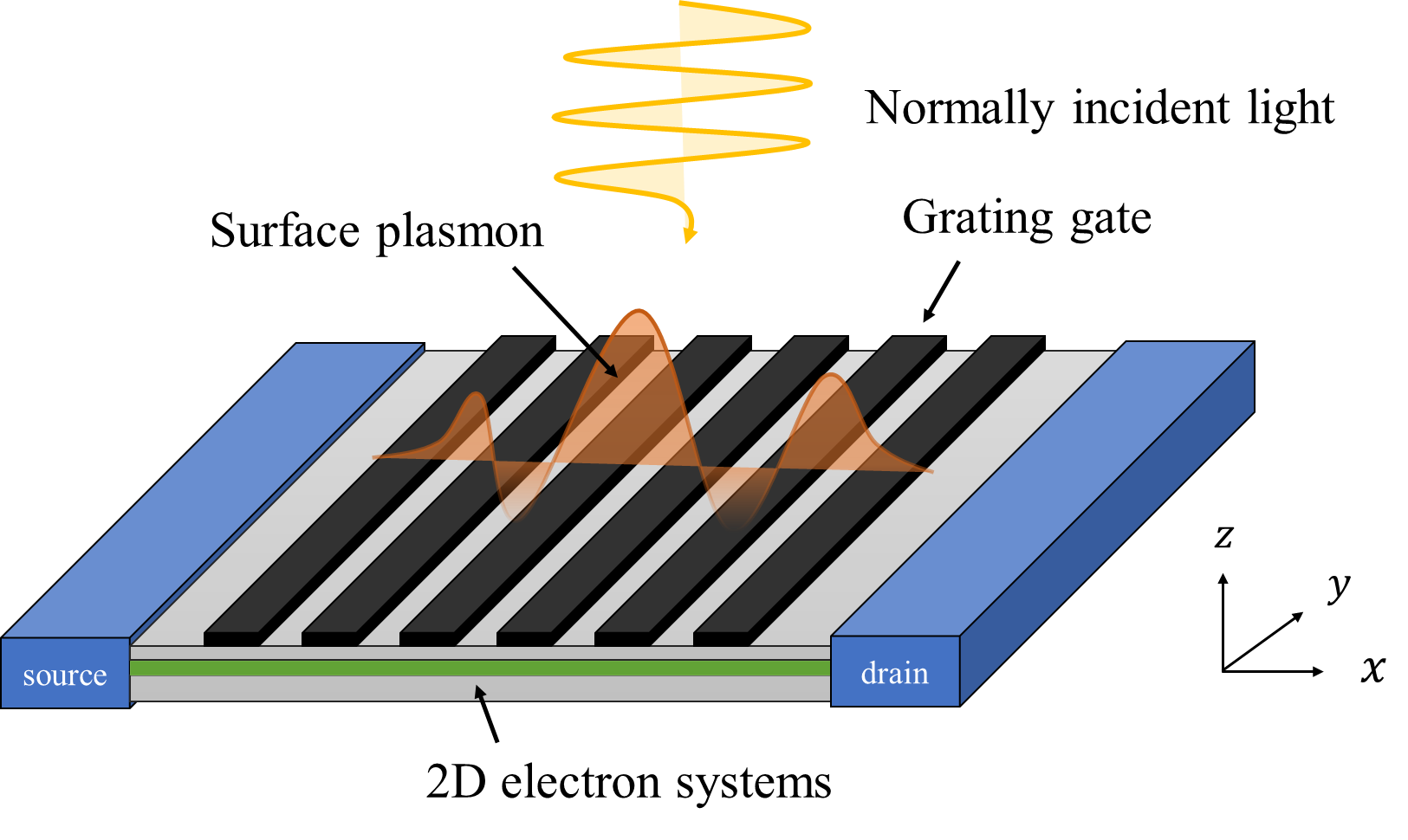}
    \caption{Our setup for plasmonically-driven geometrical photocurrent in noncentrosymmetric layerded systems with periodic grating gate.
    Other types of experimental setups, such as plasmonic cavity or antenna, would be relevant to this work.}
    \label{fig:setup}
\end{figure}

\textit{Setups.---}
Let us specify our model to describe noncentrosymmetric layered systems with plasmonic grating gates (see Fig.~\ref{fig:setup}). We assume that the grating gate spatially modulates the normally incident light $\E_0(t)=\Re[\tilde{\E}_{0}e^{i\omega t}]$, leading to the spatially dispersive electric field in 2D electron systems~\cite{Rozhansky2015,Monch2022}:
\begin{equation}
\label{eq:grating}
    \E_{in}(t,x) = \left[1 + \hat{h}\cos(qx+\phi)\right]\E_0(t),
\end{equation}
where the diagonal matrix $\hat{h}=\text{diag}\{h_x,h_y\}$ is a phenomenological parameter to determine the direction of the modulated electric field~\cite{Comment2}. Especially when $h_y$ is finite, the Faraday's law results in the presence of an out-of-plane magnetic field,
\begin{equation}
\label{eq:magnetic field}
    B(t,x)= \frac{qh_y}{\omega} \sin(qx+\phi)\Re[-i\tilde{E}_{0y}e^{i\omega t}].
\end{equation}
Since the grating gate strongly confines the incident light into the $x$-$y$ plane with a fixed small wavelength, this magnetic field has non-negligible contributions especially in the low frequency limit, leading to a novel mechanism of the photovoltaic effect as discussed below.

When the gate electrode is separated from the channel by an insulator thin film with thickness $d$ and gate-to-channel capacity $C=\varepsilon/4\pi d$, the 2D electron concentration $n(\rr,t)$ is approximately determined by the local gate-to-channel voltage $U(\rr,t)$: $n(\rr,t)= \frac{C}{e}U(\rr,t)$.
Such an approximation is often referred to as a gradual channel approximation~\cite{Shur1987,Dyakonov1993}, which is valid for smooth perturbation with $qd\ll 1$.
In summary, the total electric field is given by the sum of the incident light $\E_{in}$ and the field coming from the density perturbation: $\E=\E_{in} +(e/C)\grad n$.

Next let us consider the dynamics of electron fluids in noncentrosymmetric crystals with time-reversal symmetry.
In this paper, we focus on the hydrodynamic regime, where the rate of electron-electron scatterings $1/\tau_{e}$ exceeds that of other momentum-relaxing scatterings $1/\tau$, and thereby the total electron momentum can be regarded as a long-lived quantity~\cite{Landau_Fluid,Narozhny2017,Lucas2018_review,Narozhny2019,Polini2019}. 
Under these conditions, the electron dynamics is described by an emergent hydrodynamic theory, whose form crucially depends on the symmetry of the systems~\cite{Gorbar2018, Cook2019, Link2020,Rao2020,Varnavides2020,Varnavides2020_2,Toshio2020,Funaki2021_2,Tavakol2021,Robredo2021,Hasdeo2021,Sano2021,Wang2021,Rao2021,Varnavides2022,Varnavides2022_2,Huang2022,Friedman2022,Valentinis2022}. 
As also mentioned later, such a hydrodynamic behavior of electrons have been observed in transport experiments in various materials, and attracting a lot of interest in the last few years~\cite{Polini2019,Narozhny2022}.

For noncentrosymmetric electron fluids with parabolic dispersion near some valley $\alpha$, the formulation of electron hydrodynamics is obtained in Ref.~\cite{Toshio2020,Sano2021}. Under some reasonable approximations~\cite{comment}, 
we can transform the hydrodynamic equations for our analysis as follows:
\begin{equation}
\begin{aligned}  
\label{eq:hydrodynamic eq}
	&\pdv{\uu}{t}+(\uu \cdot \grad)\uu +\frac{\grad P}{mn} + \frac{e}{m}(\E+\uu\times \B)\\
	& \hspace{9em}+ \frac{\vb*{M}}{n} \left(\pdv{B}{t}\right)
 =-\frac{\uu}{\tau},\\
  &\pdv{n}{t}+\div(n\vb*{u})=0,
\end{aligned}
\end{equation}
where $n$ and $\rho$ are the density of particles and mass, $B$ is an applied static magnetic field, $\uu$ is the velocity field of the electron fluid, and $P$ is the pressure.
The last term on the left-hand side in the first equation is first derived in Ref~\cite{Sano2021}, and represents a geometrical anomalous force due to oscillating magnetic fields, which is closely related with the so-called gyrotropic magnetic effect~\cite{Ma2015,Zhong2016,Flicker2018,Wang2020}. Here $\vb*{M}$ is a geometrical pseudovector, and defined as the dipole component of orbital magnetic moments of Bloch electrons in the momentum space:
\begin{equation}
\vb*{M}=\sum_\alpha \vb*{M}^\alpha,\ \ \  M_{i}^\alpha \equiv \int [d\pp]\pdv{m_z^\alpha}{p_i} f_{0\alpha},
\end{equation}
where $f_{0\alpha}= [1+e^{-\beta(\e_\alpha(\pp)-\mu )}]^{-1}$ is the Fermi distribution function at the valley $\alpha$, $m_z^\alpha(\pp)$ is the orbital magnetic moment of the Bloch wavepackets~\cite{Xiao2010}, and we have introduced the notation $\int [d\pp]\equiv\int d\pp/(2\pi\hbar)^d$~\cite{definition}. 

Here it is notable that the velocity field itself is not an observable quantity.
We have to relate the velocity field $\uu$ with the observable electric current as follows~\cite{Toshio2020,Sano2021}:
\begin{equation}
\label{eq:current_general}
  \vb*{j}=-en\vb*{u}-\frac{me^2}{\hbar}(\vb*{D}\cdot\uu+YB)\cdot(\E\times \hat{\vb{e}}_z)+\cdots,
\end{equation}
where $\vb*{D}$ is another geometric pseudovector, which is often refered to as the Berry curvature dipole (BCD)~\cite{Sodemann2015}, defined as,
\begin{equation}
D_{i} = \sum_\alpha D_{i}^\alpha ,\ \  D_{i}^\alpha \equiv \int [d\pp] \pdv{\Omega_{\alpha,z}}{p_i} f_{0\alpha},
\end{equation}
and $Y$ is a geometrical scalar coefficient,
\begin{equation}
    Y = \sum_\alpha Y^\alpha,\ \  Y^\alpha=-\frac1m  \int [d\pp]\Omega_z^\alpha m_z^\alpha \partial_\epsilon f_0(\epsilon_0).
\end{equation}
$\Omega_z^\alpha(\pp)$ is the Berry curvature of
Bloch electrons in the valley $\alpha$~\cite{Xiao2010}.
Here, we note that ``$\cdots$" in Eq.~\eqref{eq:current_general} denotes the rotational currents, which causes several remarkable phenomena such as voriticity-induced anomalous current~\cite{Toshio2020}, but does not contribute to the analysis in this work.
Most important is that crystal symmetry imposes strong restrictions on the geometical pseudovectors $\vb*{M}$ and $\vb*{D}$, and thus we have to break any rotational symmetry about the $z$-axis and reduce the number of in-plane mirror lines to be less than two for these vectors to be finite.

\textit{Plasmonic QNLH effect.---} 
Here we demonstrate that the spatial modulation by grating gates gives rise to the plasmonic enhancement of the QNLH effect.
Interestingly, Ref.~\cite{Zhang2021} suggests recently that the QNLH effect has great potential for a broadband long-wavelength photodetector with small noise-equivalent power and remarkably high intenal responsivity, which is defined as the gain per absorbed power, in a broad range of frequency.
However, since its spectrum has a Lorentzian shape located at $\omega=0$ with the half width $1/\tau$, 
its external responsivity, i.e. its gain per incident power, rapidly decreases as $\omega^{-2}$ at frequencies $\omega\gg 1/\tau$, while the internal responsivity keeps a good value independent on the frequencies. Therefore, it is still an open problem how to improve the external responsivity of the QNLH effect at moderately high frequencies and whether its internal responsivity remains intact even in plasmonic resonances or not. 
In what follows, we reveal that plasmonic resonance dramatically improves the external responsivity (or the nonlinear susceptibility) by several orders of magnitude in a broad regime of frequency over $1/\tau$.

By performing a simple second-order perturbative analysis, we can easily solve the hydrodynamic equations~\eqref{eq:hydrodynamic eq} and obtain the total photocurrent, which can be decomposed into two components coming from different novel mechanisms as follows (for the detailed derivation and expression, see the supplemental materials):
\begin{equation}
\label{eq:total photocurrent}
    \vb*{j}_{DC} = \vb*{j}_{DC}^{BCD} + \vb*{j}_{DC}^{MPP},
\end{equation}
where $\vb*{j}_{DC}^{BCD}$ is a photocurrent originating from the BCD vector $\vb*{D}$, which is understood as a plasmonic version of the so-called QNLH effect~\cite{Sodemann2015}. On the other hand, 
$\vb*{j}_{DC}^{MPP}$ is another novel type of geometrical photocurrent, which comes from several nonlinear terms in Eq.~\eqref{eq:hydrodynamic eq}, such as the inertia term $(\uu\cdot \grad)\uu$.
As discussed later in more detail, since $\vb*{j}_{DC}^{MPP}$ is induced by an external oscillating magnetic field in Eq.~\eqref{eq:magnetic field}, we will hereafter refer to this contribution as {\it magnetically-driven plasmonic photogalvanic (MPP) effect}.

\begin{figure}
    \centering
      \centering
  \begin{tabular}{ll}
  (a)  &(b) \\
  \includegraphics[width=4.2cm]{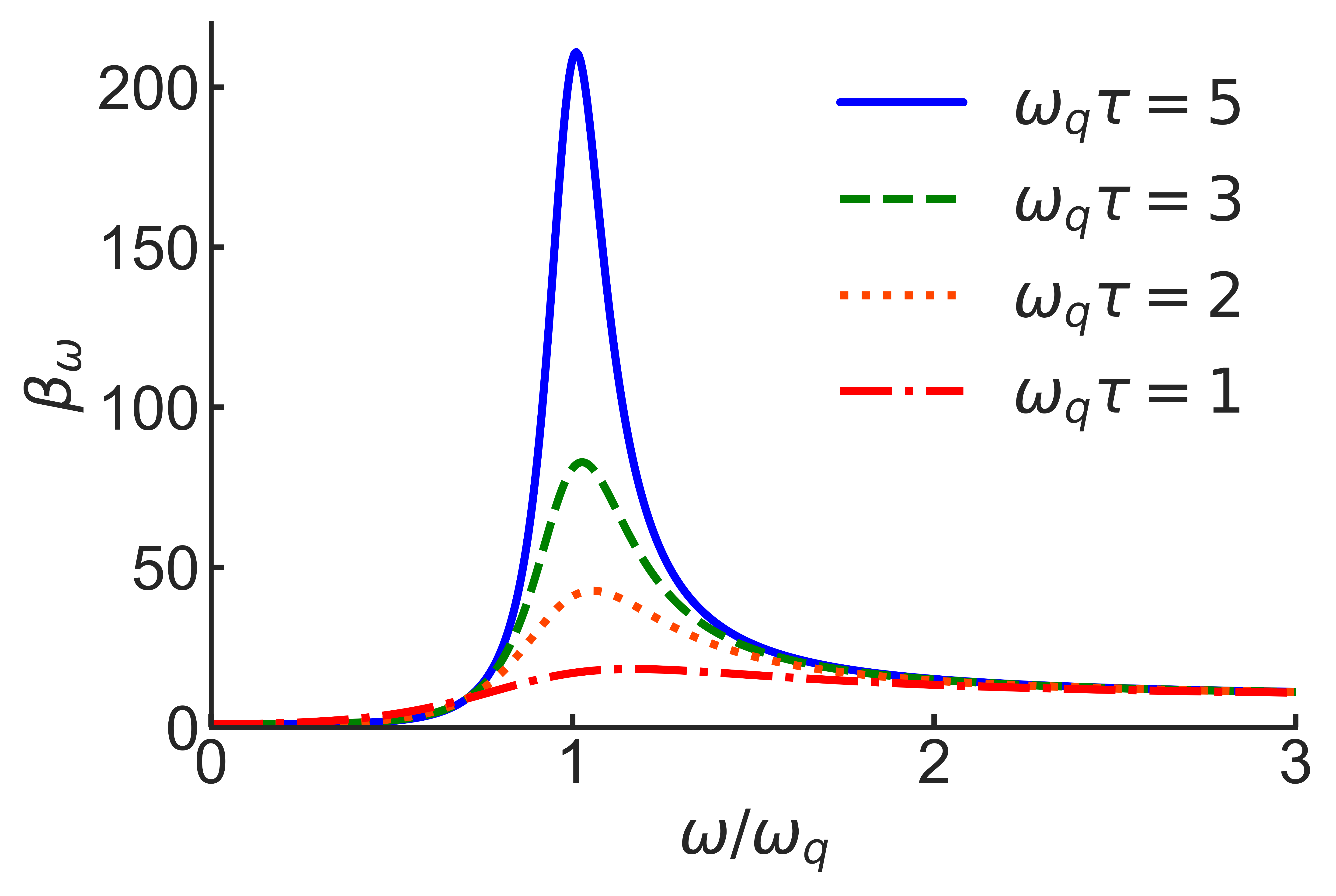}&
  \includegraphics[width=4.2cm]{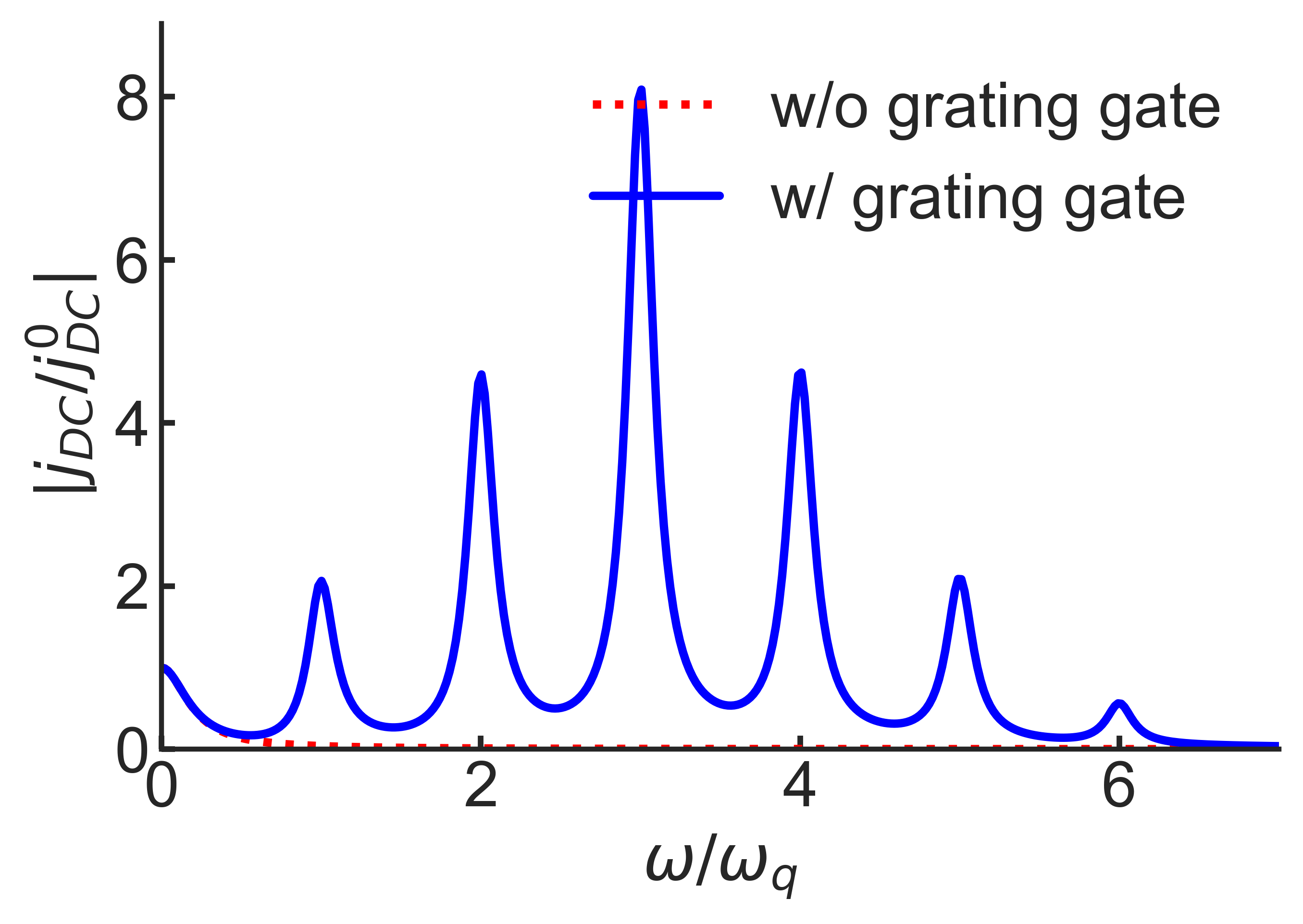}\\
  (c) & (d) \\
  \includegraphics[width=4.2cm]{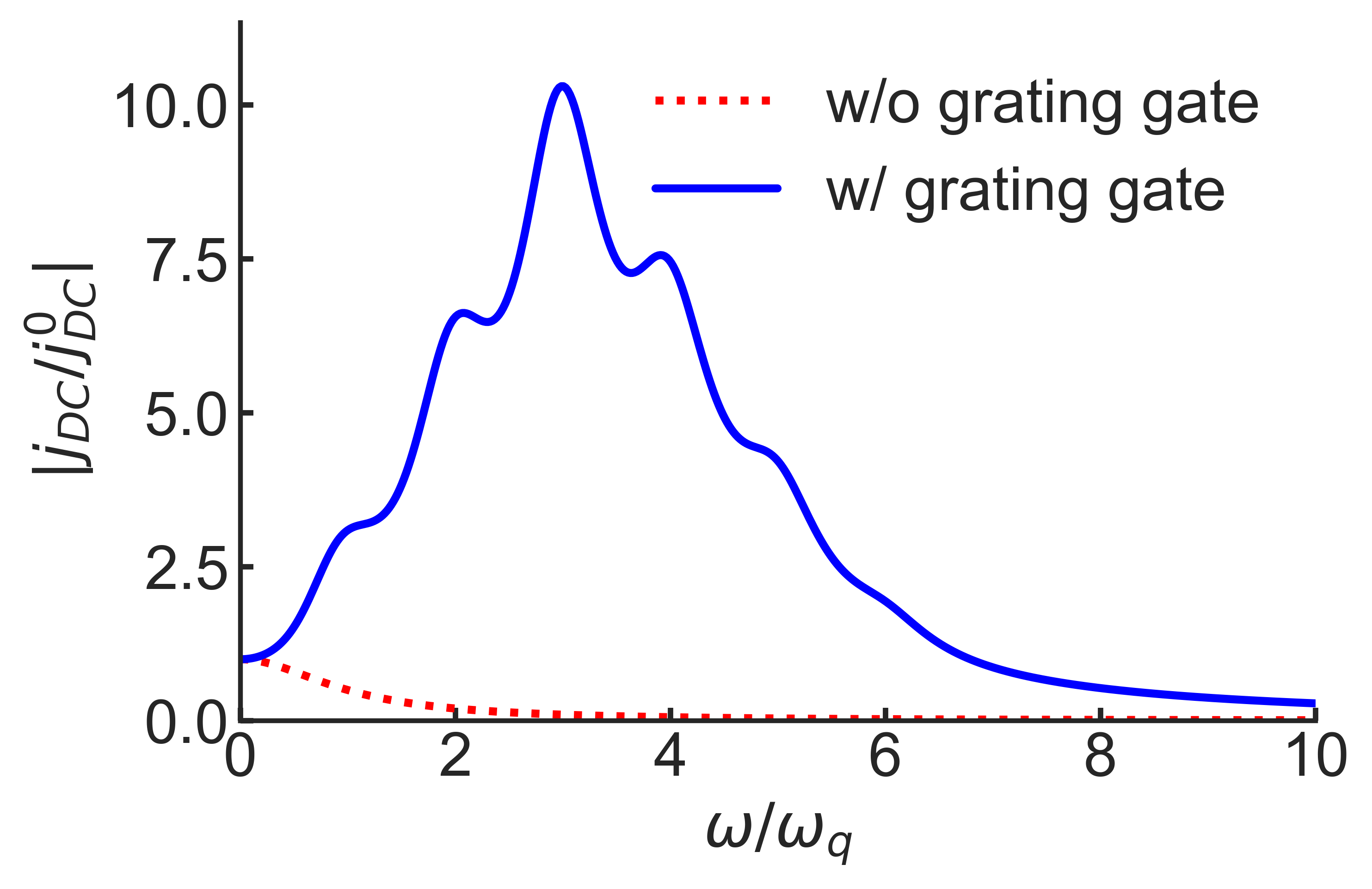}&
  \includegraphics[width=4.2cm]{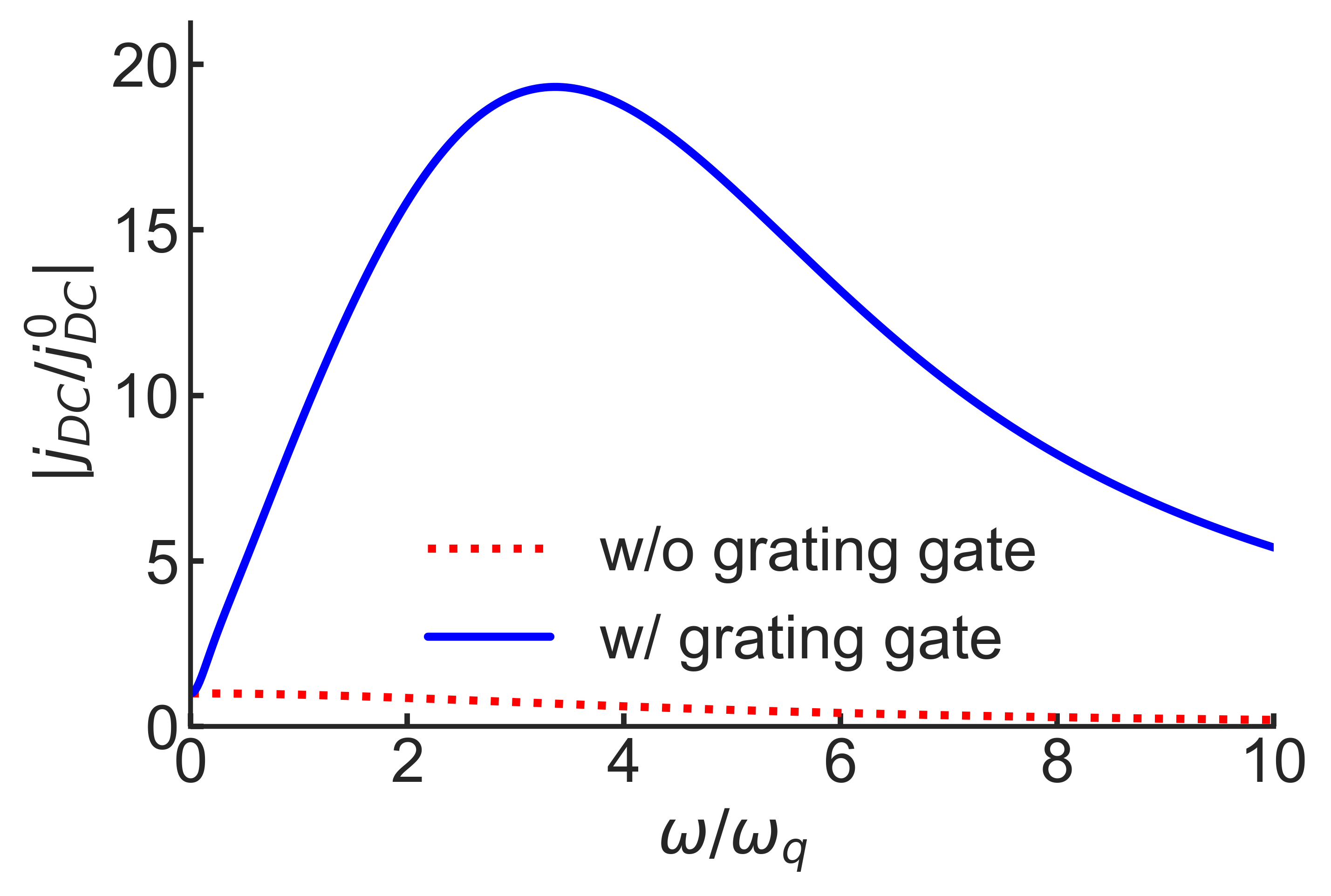}\\
  \end{tabular}
    \caption{Frequency dependence of the enhancement factor $\beta_\omega$ and the plasmonic QNLH current $j_{DC}=j_{DC}^{BCD}$. (a) We plot the enhancement factor, Eq.~\eqref{eq:enhancement factor}, which comes from only one plasmonic peak, with $h_x=4$ and various values of $\tilde{\tau}=\omega_q\tau$. (b-d) Considering the enhancement factor, Eq.~\eqref{eq:enhancement_factor_higher}, due to high-harmonic plasmons, we plot the plasmonic QNLH current, Eq.~\eqref{eq:QNLH}, normalized by $j_{DC}^0\equiv j_{DC}(\omega=0)$. We set the paramter $\tilde{\tau}=\omega_q\tau$ as (b) $\omega_q\tau =5$, (c) $\omega_q\tau =1$, and (d) $\omega_q\tau =0.2$ and, for demonstration purposes, assume phenomenologically that $(h_x^{(1)},h_x^{(2)},h_x^{(3)},h_x^{(4)},h_x^{(5)},h_x^{(6)})=(2,3,4,3,2,1)$ and $h^{(i)}=0$ ($i\geq 7$). For comparison, we also plot the spectrum of the usual QNLH current with a red dotted line.}
    \label{fig:QNLH}
\end{figure}

Here let us consider $x$-polarized incident light $\tilde{\E}_0=(\tilde{E}_{0x},0,0)$. In this case, the total photocurrent is exactly attributed only to the contribution of the QNLH term $\vb*{j}_{DC}^{BCD}$ and described by a simple beautiful form

\begin{equation}
\label{eq:QNLH}
\begin{aligned}
       \vb*{j}_{DC} = \vb*{j}_{DC}^{BCD} =-\frac{e^3}{2\hbar}
  \frac{\tau \beta_\omega }{1+(\omega\tau)^2} D_x|\tilde{E}_{0x}|^2 \hat{\vb*{e}}_y,
\end{aligned}
\end{equation}
where $\hat{\vb*{e}}_y$ is a unit vector in the $y$-direction, $\beta_\omega$ is an amplification factor due to the plasmonic resonance,
\begin{equation}
\label{eq:enhancement factor}
      \beta_\omega=
  1 + \frac{\tilde{\omega}^2(1+\tilde{\tau}^2\tilde{\omega}^2)h_x^2}{2[\tilde{\tau}^2(\tilde{\omega}^2-1)^2+\tilde{\omega}^2]},
\end{equation}
and we have introduced two dimensionless paramters: $\tilde{\tau}=\omega_q\tau$ and $\tilde{\omega}=\omega/\omega_q$. Here $\omega_q$ is the plasmon frequency $\omega_q=sq$, and $s$ is the group velocity of the plasmon.
This is one of our main results in this work, and we refer to it as the {\it plasmonic QNLH effect}.
In Fig.~\ref{fig:QNLH} (a), we have plotted the spectrum of $\beta_\omega$ for various values of $\tilde{\tau}$.


In the low-frequency limit ($\tilde{\omega}\to0$), the amplification factor $\beta_\omega$ approaches one and Eq.~\eqref{eq:QNLH} becomes equivalent to that in Ref.~\cite{Sodemann2015}, which means that the original peak of the QNLH effect at $\omega=0$ remains intact regardless of the existence of the grating gate.
On the other hand, at the plasmon frequency $\omega=\omega_q$, it features another sharp peak with a width $\Delta \omega\sim1/\tau$ in the resonant regime ($\tilde{\tau}\gg 1$), and the amplitude of the QNLH current is strongly enhanced by the dimensionless factor $|\beta_{\omega_q}|\sim|h_x\tilde{\tau}|^2/2$, compared to the case without grating gate. In particular, by utilizing near-field enhancement of gold grating gates, it is possible for the grating factor $h_x$ to be compareble to or much larger than one ($h_x\gg1$) ~\cite{Popov2015,Olbrich2016,Mass2019}. This means that the QNLH current could be enhanced by several orders of magnitude under the resonant condition ($\tilde{\tau}\gg 1$). 

In the discussion so far, we have focused on a specific harmonic mode with the wavenumber $q$ in Eq.~\eqref{eq:grating} for simplicity. However, we note that, in general, the grating gate creates high-harmonic modulations of in-plane electric fields with the wavenumbers $2q$, $3q$, $\cdots$, $Nq$~\cite{Povov2003,Fateev2010,Olbrich2011,Ivchenko2014}, which can be described as
\begin{equation}
    E_{in,i}(t,x) = \left[1 + \sum_{m=1}^\infty \hat{h}_i^{(m)}\cos(mqx+\phi_i^{(m)})\right]E_{0i}(t).
\end{equation}
These modulations result in multiple plasmonic resonant peaks at $\omega=\omega_{2q}$, $\omega_{3q}$, $\cdots$, $\omega_{Nq}$, leading to a remarkably broadband photocurrent spectrum. In particular, since the result in Eq.~\eqref{eq:QNLH} does not depend on the phases $\phi_i^{(m)}$ and the signs of $h_x^{(m)}$, all the contribution of high-harmonic plasmons flow in the same direction, and thereby strongly enhance the total photocurrent. 
This is in sharp contrast to the case of the so-called ratchet effect~\cite{Olbrich2011,Rozhansky2015}, which is strongly depend on these parameters, and thus, each plasma mode often cancels each other.
In conclusion, enhancement factor~\eqref{eq:enhancement factor} is modified by high-harmonic plasmons as follows:
\begin{equation}
\label{eq:enhancement_factor_higher}
\beta_\omega = 1 + \sum_{m=1}^\infty \frac{\tilde{\omega}^2(1+\tilde{\tau}^2\tilde{\omega}^2)(h_x^{(m)})^2}{2[m^2\tilde{\tau}^2(\tilde{\omega}^2-m^2)^2+\tilde{\omega}^2]}.
\end{equation}
In Fig.~\ref{fig:QNLH} (b-d), We have plotted the spectrum of the plasmonic QNLH current (blue line), and campared it with that of the normal QNLH effect.
From these figures, we find that the QNLH current is dramatically enhanced by several orders of amplitude, over a very broad range of frequency above the original frequency threshold $1/\tau$.
Similar enhancement effects due to high-harmonic plasma modes have ever been discussed in the context of terahertz light absorption~\cite{Povov2003,Muravjov2010,Guo2013} and plasmonic ratchet effect~\cite{Popov2015,Olbrich2016}.

\textit{Universal internal responsivity.---} 
Here, assuming more general situations, we elucidate a universal relation between the photocurrent induced by the BCD and the light absorption by 2D electron systems.
First, in general, the BCD-induced photocurrent is obtained from Eq.~\eqref{eq:current_general} in the following form,
\begin{equation}
    \vb*{j}_{DC}^{BCD} = -\frac{me^2}{\hbar}
    \expval{(\vb*{D}\cdot\vb*{u}(\rr,t))(\E(\rr,t)\times \hat{\vb*{e}}_z)}_{t,\rr},
\end{equation}
where $\expval{\cdots}_{t,\rr}$ denotes the time and space averaging over the periods.
Especially for $x$-polarized incident light, it leads to 
$\vb*{j}_{DC}^{BCD} = \frac{me^2}{\hbar}D_x
\expval{u_xE_x}_{t,\rr} \hat{\vb*{e}}_y$. On the other hand, the optical power absorbed by 2D electron systems can be calculated as $\mathcal{P}=S\expval{j_xE_x}_{t,\rr}$, where $S=L_xL_y$ is the area of our system and $L_i$ is the sample's size in the $i$-direction. In the linear order of external perturbations, the electric current is related with the velocity field as $j_x=-en_0u_x$ from Eq.~\eqref{eq:current_general}, where $n_0$ is the equilibrium particle density. Combining these formulas, we reach the desired universal relation between $\vb*{j}_{DC}^{BCD}$ and $\mathcal{P}$ as follows: 
\begin{equation}
\label{eq:universal relation}
    \vb*{j}_{DC}^{BCD} = - \frac{emD_x\mathcal{P}}{\hbar n_0 S} \hat{\vb*{e}}_y.
\end{equation}
This relation means that the plasmonic QNLH effect discussed above comes from the plasmonic enhancement of the total optical absorption by grating gate.
As understood from the derivation, Eq.~\eqref{eq:universal relation} will be satisfied in more generic situations beyond our 2D grating model, such as plasmonic cavities~\cite{Li2017_2,Hugall2018,Xiao2018,Epstein2020} or antennas~\cite{Bandurin2018_2,Ullah2020}, as far as the frequency is low enough for interband transitions to be negligible~\cite{Note}. Here we note that, as easily checked from the formula, 
Eq.~\eqref{eq:universal relation} does not hold for the circular photogalvanic current induced by the BCD~\cite{comment}. This means that higher efficiency could be achieved for circularly polarized light, which is analogous to recent proposals in Ref.~\cite{Onishi2022,Shi2022}.

From Eq.~\eqref{eq:universal relation}, we can immediately obtain the internal current responsivity of the BCD-induced Hall photocurrent, which is one of the most important
figures of merit quantifying the performance of THz detectors~\cite{Koppens2014} and defined as the current gain per absorbed light power,
\begin{equation}
\label{eq:responsivity}
    \mathcal{R}_I \equiv \frac{|I_y|}{\mathcal{P}}
    =\left|\frac{emD_x}{\hbar n_0 L_y}\right|.\ \ \ \ \  (I_y=j_{DC,y}^{BCD}L_x)
\end{equation}
This is another important result in this paper. Eq.~\eqref{eq:responsivity} states that the responsivity is entirely determined by the band structure (and the carrier density) of electron sytems, and completely independent on incident frequencies and their environment such as grating or cavity structures.
Clearly, this property is very beneficial for computational material design toword terahertz-infrared photodetectors. To realize a high-performance photodetector utilizing the BCD-induced photocurrent, first we should search quantum materials with a collossal effective mass $m$ and BCD by performing ab-initio calculations or experiments, and then, improve their optical absorption by decorating or designing those promising materials with some plasmonic or cavity structures.

For the latter purpose, 2D layered materials, which are very flexible to various device designs, seem to be more advantageous than 3D bulk materials. Recent experiments~\cite{Ma2019} have reported that the external voltage responsivity of bilayer WTe${}_2$ reaches a value of $2\times10^4$ V/W${}^{-1}$ around $\omega\simeq 100$ Hz at $T=10$ K, which is notably large and comparable to the best values in existing rectifiers~\cite{Ma2019,Auton2016}.
Furthermore, Ref.~\cite{Zhang2022} has theoretically suggested that strained twisted bilayer graphene achieves a further large responsivity that is 20 times larger than the above values. However, since these materials work well only at low temperature, further investigations of promising materials, which show a remarkably large value of the BCD, will be needed to realize terahertz photodetectors working at room temperature.

\textit{Magnetically-driven plasmonic photogalvanic effect.---}
Next let us consider a novel type of photocurrent, $\vb*{j}_{DC}^{MPP}$, obtained in Eq.~\eqref{eq:total photocurrent}, which is regarded as a spatially dispersive correction to the total photocurrent and proportional to $q^2$ or $q^4$. 
For this reason, this effect is peculiar to spatially structured systems like our grating model, and does not appear in spatially uniform cases.

As shown in detail in the supplemental materials, the MPP effect originates from an anomalous driving force induced by oscillating magnetic fields ($\propto \vb*{M} (\partial B/\partial t)$) in Eq.~\eqref{eq:hydrodynamic eq}, and thus, they are described by the geometrical pseudovector, $\vb*{M}$, i.e., the dipole moment of orbital magnetic moments of Bloch electrons in the momentum space (for the detailed derivation and expression, see the supplemental materials). 
In particular, at plasmon frequencies, the MPP current also has a sharp peak, as in the case of the plasmonic QNLH effect, and the peak amplitude is obtained under the resonant condition ($\tilde{\tau}\gg1$) as follows~\cite{neglectY}:
\begin{equation}
\begin{aligned}
    \vb*{j}_{DC}^{MPP}(\omega=\omega_q) =& \frac{e^2\tau}{4ms^2}[
    \tilde{\tau} h_xh_y\mathcal{F}_z(\vb*{M}+2M_x\hat{\vb*{e}}_x)\\
    & \quad\quad +\tilde{\tau}^2 h_xh_y\mathcal{L}_{xy} M_x\hat{\vb*{e}}_x] + \mathcal{O}(\tilde{\tau}^0).
\end{aligned}
\end{equation}
Here we have introduced $\mathcal{F}_{z}=\frac{i}{2}(\tilde{E}_{0x}\tilde{E}_{0y}^* - \tilde{E}_{0y}\tilde{E}_{0x}^*)$ and $\mathcal{L}_{xy} =\frac12 (\tilde{E}_{0x}\tilde{E}_{0y}^* + \tilde{E}_{0y}\tilde{E}_{0x}^*)$, each of which represents a circular photogalvanic effect and a linear photogalvanic effect.
Focusing on its circular photogalvanic effect in the $x$-direction, the value of MPP current is around 0.01 nA/W with typical values of parameters, $m\sim m_e$, $s\sim 1\times10^6$ m/s, $\tau\sim 1\times 10^{-12}$ s and an estimated value of $M_x$ obtained in Ref.~\cite{Sano2021} for strained graphene, $M_x\sim 3\times 10^{17}$ s$\cdot$A/kg$\cdot$m, assuming the resonant case $|\tilde{\tau}h_xh_y|\gtrsim 10$.
Although this is much smaller than the measured value of QNLH current ($\sim 100$ nA/W) in monolayer WTe${}_2$~\cite{Xu2018} around $\omega\simeq 30$ THz at 150 K, we might be able to improve the MPP current furthermore by seeking materials with a much larger value of $\vb*{M}$.
In such a situation, since the plasmonic term of the BCD-induced circular photocurrent is proportional to $\omega^2-\omega_q^2$ and thus vanishes at the plasmon frequency, the MPP effect will dominate the total photocurrent. This might be one of good optical probes for the geometical strctures of Bloch electrons in 2D quantum systems.

\textit{Discussion.---}
Here we briefly discuss possible candidates to observe the novel types of plasmonic photocurrents obtained in this work.
In the past few years, many pieces of evidence for hydrodynamic electron flow have been reported in various materials, including monolayer/bilayer graphene~\cite{Bandurin2016, Crossno2016, Kumar2017, Bandurin2018,  Sulpizio2019, Berdyugin2019}, GaAs quantum wells~\cite{Molenkamp1994,Molenkamp1995, Braem2018, Gusev2018-2, Gusev2018, Levin2018}, 2D monovalent layered metal PdCoO${}_2$~\cite{Moll2016}, Weyl semimetal WP${}_2$~\cite{Gooth2018}, and WTe${}_2$~\cite{Vool2021,Choi2022,Steinberg2022}.
Among these materials, promising candidates for our work are graphene with some deformation and layered transition metal dichalcogenide WTe${}_2$. These materials have crystal symmetries low enough to exhibit intriguing optical phenomena, such as the QNLH effect~\cite{Sodemann2015}, which is required for the geometrical pseudovectors $\vb*{D}$ and $\vb*{M}$ to be finite. As a matter of fact, the QNLH effect itself has already been observed in layered WTe${}_2$~\cite{Xu2018,Ma2019,Wang2019,Kang2019} and artificially corrugated bilayer graphene~\cite{Ho2021}. In particular, bilayer WTe${}_2$ is reported to show remarkably high responsivity~\cite{Ma2019} as already mentioned, and further dramatic enhancement of the BCD is suggested by twisting the two layers in Ref.~\cite{He2021}.

Another possible candidate is (110) quantum well in GaAs, since it also has crystal symmetries low enough to show the QNLH effect~\cite{Moore2010,Ganichev2001,Diehl2007,Olbrich2009} and another type of GaAs quantum well has already shown various hydrodynamic signatures~\cite{Molenkamp1994,Molenkamp1995, Braem2018, Gusev2018-2, Gusev2018, Levin2018}.
Furthermore, twisted bilayer graphene, a novel layered system attracting great interest recently, might also be a candidate for our work, since this material is theoretically suggested to realize the hydrodynamic regime~\cite{Zarenia2020} and to show a remarkably high responsivity of the QNLH effect~\cite{Zhang2022}.

Finally, we give a brief discussion about the viscosity effect on our results~\cite{comment}. In the context of electron hydrodynamics, viscosity is regarded as a key ingredient to characterize electron dynamics in the hydrodynamic regime. Actually, a lot of recent experiments have ever been devoted to measurements of the signature of viscosity in nonlocal transport phenomena~\cite{Bandurin2016,Moll2016,Gooth2018,Levin2018,Gusev2018,Gusev2018-2,Berdyugin2019}. 
By turning on viscosity term phenomenologically in Eq.~\eqref{eq:hydrodynamic eq}, we find that the width of plasmonic peaks in Eq.~\eqref{eq:QNLH} is modified from $1/\tau$ to $1/\tau+(\nu+\zeta) q^2$, where $\nu$ and $\zeta$ are the kinetic viscosity and the bulk viscosity. This means that, for typical values of parameters $\tau=1\times10^{-12}\ [\mathrm{s}^{-1}]$ and $\nu=1\times10^{-1} \ [\mathrm{m^2s^{-1}}]$~\cite{Bandurin2016}, viscosity causes non-negligible contributions to the plasmon lifetime when the cycle length $L=2\pi/q$ becomes $\mu$m-order or less. Consequently, it might be possible to optically probe mysterious aspects of strongly correlated electron systems such as twisted bilayer graphene~\cite{Cao2020,Gonzalez2020,Cha2021,Sarma2022}, through the peculiar temperature dependence of plasmon lifetime, since the electron viscosity behaves as $\nu\sim v_Fl_{ee}\propto 1/T^2$ in Fermi liquids~\cite{Qian2005,Alekseev2016,Gooth2018}, while $\nu \propto 1/T$ in typical non-Fermi liquids~\cite{Davison2014,Gooth2018,Narozhny2019,Kovtun2005}.

\textit{Conclusion.---}
In summary, based on an electron hydrodynamic theory, we have formulated plasmonically-driven  geometrical photocurrents in nocentrosymmetric 2D layered systems with periodic grating gates. Our framework can be generalized to various types of problems in plasmonics, such as plasmonic responses of 1D vdW materials~\cite{Freitag2013,Guo2022} and gate-controlled optical activity~\cite{Kim2017}, plasmon-to-current converters~\cite{Povov2013,Kojori2016}. This provide us with a new way to investigate the role of quantum geometry in plasmonics, leading to a promising route toward a novel type of highly sensitive, broadband and electrically-controllable terahertz plasmonic devices.

\textit{Acknowledgments.---}
We would like to express our 
special thanks to Hikaru Watanabe and Koichiro Tanaka for useful comments. We are also grateful to Gen Tatara, Hiroshi Funaki, Ryotaro Sano, and Akito Daido for valuable discussions. This work is partly supported by JSPS KAKENHI (Grants JP20J22612, JP18H01140 and JP19H01838).

\bibliographystyle{apsrev4-1}
\bibliography{citation}

\clearpage

\renewcommand{\thesection}{S\arabic{section}}
\setcounter{section}{0}
\renewcommand{\theequation}{S\arabic{equation}}
\setcounter{equation}{0}
\renewcommand{\thefigure}{S\arabic{figure}}
\setcounter{figure}{0}

\onecolumngrid
\begin{center}
{\large {\bfseries Supplemental Materials for \\ ``Plasmonic quantum nonlinear Hall effect in noncentrosymmetric 2D materials"}}
\end{center}

\subsection{I. Formulation}

First we give some supplemental information about our theoretical models and some approximations applied to them in our work. In general, the continuity equations for electron momentum and particle density are obtained in the relaxation-time approximation as follows~\cite{Toshio2020,Sano2021}:
\begin{align}
\label{eq:continue}
  &\partial_tN+\div\vb*{J}^n=0,\\
  &\partial_tP_{i}+\partial_j\Pi_{ij}=F_i+\Gamma_i,
\end{align}
where $N$ and $P_i$ are respectively the density of electron momentum and particle number, $\vb*{J}$ and $\Pi_{ij}$ are the flux of them, $\vb*{F}$ is the external force due to the applied electromagnetic fields, $\vb*{\Gamma}$ is the disipative force due to the momentum relaxing scatterings.
As formulated in Ref.~\cite{Toshio2020,Sano2021}, for  two-dimensional (2D) noncentrosymmetric electron fluids with parabolic dispersion near some valleys, these values are related with the hydrodynamic variables, such as velocity fields $\uu$, as
\begin{align}
\label{eq:flux}
  N &= n+\frac{meB}{\hbar}(\vb*{D}+\vb*{N})\cdot \uu+\mathcal{O}(B^2),\quad\quad
  \vb*{P} = mn\uu+\frac{eB}{\hbar}(\vb*{C}+\vb*{M}),\\
  \vb*{J}^n&=n\vb*{u}+\frac{me}{\hbar}(\vb*{D}\cdot\uu+YB)\cdot(\E\times \hat{\vb{e}}_z),\\
  \Pi_{ij}&=mnu_iu_j+p\delta_{ij}+\frac{e}{\hbar}\epsilon_{jk}E_kC_{i}+mB\left[( \vb*{M}\cdot\uu)\delta_{ij}+M_{i}u_j\right]+\mathcal{O}(B^2),\\
  \vb*{F}&=-en(\vb*{E}+\vb*{u}\times\vb*{B}),\quad\quad
  \vb*{\Gamma}=-mn\uu/\tau,
\end{align}
where $\E$ is an applied in-plane electric field, $\B=(0,0,B)$ is an applied out-of-plane magnetic field, $n$ is the particle density without the correction due to magnetic fields, and $\vb*{C}$, $\vb*{D}$, $\vb*{M}$, and $\vb*{N}$ are the geometrical pseudovectors originating from the Berry curvature or the magnetic moment of the electron wavepackets (For the definition of the pseudovectors, see Ref.~\cite{Sano2021} and the main text). These equations are correct up to the second order of the external perturbations. Here we note that $\mathcal{O}(B^2)$ terms in $n$ and $\Pi_{ij}$ are not explicitly shown, since they never contribute to the results in this work. For example, in the analysis of the photocurrent, they always appear in the form of the time or spatial derivative, $\partial_{t,x}({O}(B^2))$, and thus do not contribute to the spatially and temporally uniform currents. For the same reason, we do not need to consider the terms in the form $\partial_{t,x}({O}(E^2, BE, B^2))$ in the following analyses.
Furthermore, we note that, although geometrical pseudovectors, such as $\vb*{C}$ and $\vb*{D}$, depend on time $t$ and spatial position $\rr$ through chemical potential $\mu(t,\rr)$ or temperature $T(t,\rr)$, we can neglect higher-order corrections due to these dependence for the above reason.

Combining Eqs.~\eqref{eq:continue} and~\eqref{eq:flux}, we obtain the 
hydrodynamic equations for noncentrosymmetric electron fluids with parabolic dispersion:
\begin{equation}
\begin{aligned}  
	\pdv{\uu}{t}+(\uu \cdot \grad)\uu +\frac{\grad P}{mn} + \frac{e}{m}(\E+\uu\times \B)
	+ \frac{\vb*{M}}{n} \left(\pdv{B}{t}\right)+\cdots
 =-\frac{\uu}{\tau},
\end{aligned}
\end{equation}
\begin{equation}
\begin{aligned}  
  \pdv{n}{t}+&\div(n\vb*{u})+\cdots=0,
\end{aligned}
\end{equation}
where ``$\cdots$" denotes the negligible terms in the form $\partial_{t,x}({O}(E^2, BE, B^2))$.
As explained in the main text, the total electric field is described as $\E=\E_{in} +(e/C)\grad n$ in our metamaterial structure, and thus the above equations are rewritten in a more explicit form: 
\begin{equation}
\label{eq:hydro_perturbative}
\begin{aligned}  
	\pdv{\uu}{t}+(\uu \cdot \grad)\uu 
	+ \frac{e}{m}(\E_{in}+\uu\times \B) +\frac{s^2}{n_0}\grad \delta n
	+ \frac{\vb*{M}}{n_0}\left(1-\frac{\delta n}{n_0}
	\right) \left(\pdv{B}{t}\right)+\cdots
 =-\frac{\uu}{\tau},
\end{aligned}
\end{equation}
\begin{equation}
\begin{aligned}  
  \pdv{\delta n}{t}
  +n_0\grad \uu+\cdots=0,
\end{aligned}
\end{equation}
where $s^2=n_0(e^2/mC+n_v/m\nu)$ is the group velocity of the plasmon, $\nu=m/2\pi\hbar^2$ is the density of states per valley, $n_v$ is the number of valleys, $n_0$ is the uniform particle density without perturbations, and then we introduced the notation $\delta n\equiv n-n_0$. In the derivation, we have used the formula $P\simeq n_vn^2/2\nu + n_v\nu T^2\pi^2/6$ in the 2D degenerate electrons with parabolic dispersion and neglect the thermoelectrical force $F_{th}\propto \grad T^2$ since it can be estimated as $F_{th}\propto \grad{\uu^2}\propto \partial_{t,x}({O}(E^2, BE, B^2))$ discussed in Ref.~\cite{Rozhansky2015}.

We can calculate electric current density $\vb*{j}(x,t)$ by solving the above hydrodynamic equations under external fields, and substituting the obtained solution $\uu(x,t)$ and $n(x,t)$ to the formula~(4) in the main text, which relate the hydrodynamic variables and electric currents as follows:
\begin{equation}
  \vb*{j}=-en\vb*{u}-\frac{me^2}{\hbar}(\vb*{D}\cdot\uu+YB)\cdot(\E\times \hat{\vb{e}}_z)+\cdots.
\end{equation}
As noted in the main text, the last term ``$\cdots$" in Eq.~(4) denotes the rotational currents, which cause several remarkable phenomena, such as vorticity-induced anomalous current~\cite{Toshio2020}, but are negligible for the analysis in this work, since rotational components never contributes to spatially uniform current.    
The first term in Eq.~(4) is a familiar part known in conventional hydrodynamic theory, and the others are geometrical contributions due to the symmetry lowering of the fluids. In particular, the term proportional to the coefficient $\vb*{D}$ describes the contribution of the so-called quantum nonlinear Hall (QNLH) effect~\cite{Sodemann2015,Toshio2020}.

\subsection{II. Derivation of plasmonically-enhanced photocurrent}

In the following, we perform a perturbative calculation to analyze plasmonically-enhanced photocurrents in noncetrosymmetric electron fluids. For this purpose, we 
perturbatively expand the velocity field $\uu$, particle density $n$, and the total electric field $\E$ with respect to $\tilde{\vb*{E}}_0$ as follows:
\begin{equation}
    \uu = \uu_1+\uu_2+\cdots,\ \ \ \ \ n=n_0+\delta n=n_0+\delta n_1+\delta n_2+\cdots,\ \ \ \ \ 
    \E = \E_1 + \E_2+\cdots,\ \ \ \ \ 
    B = B_1.
\end{equation}
With this notation, from Eq.~(4) in the main text, the leading photocurrent can be written as a sum of several contributions:
\begin{equation}
\label{eq:photocurrent_general}
    \vb*{j}_{DC} = -e\expval{\delta n_1\uu_1 + n_0 \uu_2}_{t,x}
    -\frac{me^2}{\hbar}\expval{(\vb*{D}\cdot\uu_1+YB_1)\cdot(\E_1\times \hat{\vb{e}}_z)}_{t,x},
\end{equation}
where $\expval{\cdots}_{t,x}$ denotes the time and space averaging over the periods.

The first order contributions can be easily calculated from Eq~\eqref{eq:hydro_perturbative} and we obtain 
\begin{equation}
\begin{aligned}
\label{eq:first_order}
  u_{1,x} &=\Re\left[ 
  \frac{e^{i\omega t}}{\omega^2-\omega_q^2 -i\omega/\tau}\left(
  \frac{ie\omega}{m}(h_x\tilde{E}_{0x})\cos(qx+\phi)+
  \frac{i\omega q}{n_0}(h_y\tilde{E}_{0y})M_x\sin(qx+\phi)
  \right)
  -\frac{e}{m}\frac{e^{i\omega t}}{i\omega+1/\tau} \tilde{E}_{0x}
  \right],\\
  u_{1,y} &=\Re\left[ -
  \frac{e^{i\omega t}}{i\omega+1/\tau}\left(
  \frac{e}{m}(h_y\tilde{E}_{0y})\cos(qx+\phi)+
  \frac{q}{n_0}(h_y\tilde{E}_{0y})M_y\sin(qx+\phi)
  \right)
  -\frac{e}{m}\frac{e^{i\omega t}}{i\omega+1/\tau} \tilde{E}_{0y}
  \right],\\
  \delta n_1 &=\Re\left[ 
  \frac{e^{i\omega t}}{\omega^2-\omega_q^2 -i\omega/\tau}\left(
  \frac{en_0 q}{m}(h_x\tilde{E}_{0x})\sin(qx+\phi)-
  (h_y\tilde{E}_{0y}q^2)M_x\cos(qx+\phi)
  \right)
  \right].
\end{aligned}
\end{equation}
These results mean that the spatial modulation by the grating gate ($\vb*{h}\neq 0$) causes a plasmonic resonance denoted by the prefactor $(\omega^2-\omega_q^2 -i\omega/\tau)^{-1}$, whose imaginary part has a strong peak with the half width $1/\tau$  near the plasmon frequency $\omega\sim\omega_q$ under the resonant condition.

Next let us consider the contributions in the second order of perturbations. From Eq.~\eqref{eq:hydro_perturbative}, the second order term of the velocity fields $\uu_2$ satisfies
\begin{equation}
    \pdv{\uu_2}{t}+(\uu_1 \cdot \grad)\uu_1 
	+ \frac{e}{m}(\uu_1\times \B_1) +\frac{s^2}{n_0}\grad \delta n_2
	+ \frac{\vb*{M}}{n_0}\left(\frac{\delta n_1}{n_0}
	\right) \left(\pdv{B_1}{t}\right)+\cdots
 =-\frac{\uu_2}{\tau}.
\end{equation}
Focusing on the time and space average of the velocity fields $\uu_2$, this equation leads to the relation
\begin{equation}
    \expval{\uu_2}_{t,x}=-\tau \expval{(\uu_1 \cdot \grad)\uu_1+\frac{e}{m}(\uu_1\times \B_1)+\frac{\vb*{M}}{n_0}\left(\frac{\delta n_1}{n_0}
	\right) \left(\pdv{B_1}{t}\right)}_{t,x},
\end{equation}
and each term can be calculated as follows:
\begin{equation}
\begin{aligned}
  \expval{(\uu_1 \cdot \grad)\uu_1}_{t,x}&= \frac{e q^2}{4mn_0} \frac{\omega^2h_xh_y}{(\omega^2-\omega_q^2)^2+ (\omega/\tau)^2}\Re\left[E_{0x}E_{0y}^*
  \right](M_x\hat{\vb*{e}}_x)\\
  &\ \ \ \ \ \ +\frac{e q^2}{4mn_0}\Re\left[
  \frac{i\omega}{(\omega^2-\omega_q^2 -i\omega/\tau)(i\omega-1/\tau)}
  ((h_x\tilde{E}_{0x})(h_y\tilde{E}_{0y}^*)M_y
  +|h_y\tilde{E}_{0y}|^2M_x)
  \right]\hat{\vb*{e}}_y,\\
  \expval{\frac{e}{m}(\uu_1\times \B_1)}_{t,x}&=  \frac{e q^2}{4mn_0}
  \Re\left[
  \frac{1}{\omega^2-\omega_q^2 -i\omega/\tau} |h_y\tilde{E}_{0y}|^2
  \right] M_x\hat{\vb*{e}}_y\\
  &\ \ \ \ \ \ -\frac{e q^2}{4mn_0}
  \Re\left[
  \frac{1}{\omega^2-i\omega/\tau} |h_y\tilde{E}_{0y}|^2
  \right] M_y\hat{\vb*{e}}_x\\
  \expval{\frac{\vb*{M}}{n_0}\left(\frac{\delta n_1}{n_0}
	\right) \left(\pdv{B_1}{t}\right)}_{t,x}&= \frac{eq^2}{4mn_0}\Re\left[
	\frac{1}{\omega^2-\omega_q^2 -i\omega/\tau}
	(h_x\tilde{E}_{0x})(h_y\tilde{E}_{0y}^*)\vb*{M}
	\right],
\end{aligned}
\end{equation}
where $\hat{\vb{e}}_i$ is a unit vector in the direction of $i$-axis.

Substituting these results into Eq.~\eqref{eq:photocurrent_general}, we obtain the total photocurrent $\vb*{j}_{DC}$ as the sum of the contributions arising from different mechanisms as follows:
\begin{equation}
    \vb*{j}_{DC} =\vb*{j}_{DC}^{BCD} + \vb*{j}_{DC}^{MPP}, 
\end{equation}
where $\vb*{j}_{DC}^{BCD}$ is the BCD-induced photocurrent, 
\begin{equation}
\begin{aligned}
  \vb*{j}_{DC}^{BCD} &=\frac{e^3}{2\hbar}\Re \left[ \frac{1}{i\omega+1/\tau}
  (\vb*{D}\cdot\tilde{\E}_0)\cdot (\tilde{\E}_0^*\times \hat{\vb{e}}_z)
  +\frac12\left(- \frac{i\omega}{\omega^2-\omega_q^2 -i\omega/\tau}D_x h_x\tilde{E}_{0x}
  + \frac{1}{i\omega+1/\tau}
  D_y h_y\tilde{E}_{0y}
  \right)(\hat{h}\tilde{\E}_0^*\times \hat{\vb{e}}_z)
  \right]\\
  & \quad \quad - \frac{e^5n_0q^2}{4\hbar mC} 
  \Re \left[
  \frac{1}{(i\omega+1/\tau)(\omega^2-\omega_q^2+i\omega/\tau)}\left(
  (h_y\tilde{E}_{0y})(h_x\tilde{E}_{0x}^*) 
  +\left(\frac{mq}{en_0}\right)^2M_y|h_y\tilde{E}_{0y}|^2
  \right)
  \right] (D_y\hat{\vb{e}}_y).\\
\end{aligned}
\end{equation}
On the other hand, $\vb*{j}_{DC}^{MPP}$ is a magnetically-driven plasmonic photocurrent, which is proportional to the geometirical pseudovector $\vb*{M}$, and obtained as the sum of contributions of several nolinear terms in the hydrodynamic equations or anomalous currents proportional to the geometrical scalar $Y$,
\begin{equation}
    \vb*{j}_{DC}^{MPP} = \vb*{j}_{DC}^{density} + \vb*{j}_{DC}^{inertia} +\vb*{j}_{DC}^{Lorentz}  + \vb*{j}_{DC}^{Y},
\end{equation}
where
\begin{equation}
\begin{aligned}
  \vb*{j}_{DC}^{density} &=\frac{e^2q^2}{4m}\Re\left[
	\frac{\tau}{\omega^2-\omega_q^2 -i\omega/\tau}
	(h_x\tilde{E}_{0x})(h_y\tilde{E}_{0y}^*)\vb*{M}
	\right]\\
	&\quad \quad \quad 
	+\frac{e^2q^2}{4m} \frac{\omega}{(\omega^2-\omega_q^2)^2+ (\omega/\tau)^2} \Re\left[
	i(h_x\tilde{E}_{0x})(h_y\tilde{E}_{0y}^*)-i(h_y\tilde{E}_{0y})(h_x\tilde{E}_{0x}^*)
	\right](M_x\hat{\vb{e}}_x)\\
	&\quad \quad \quad 
	+\frac{e^2q^2}{4m} \Re\left[
	\frac{i}{(i\omega-1/\tau)(\omega^2-\omega_q^2 -i\omega/\tau)}
	\left(|h_y\tilde{E}_{0y}|^2M_x-(h_y\tilde{E}_{0x})(h_x\tilde{E}_{0y}^*)M_y
	\right)
	\right]\hat{\vb{e}}_y,\\
  \vb*{j}_{DC}^{inertia} &=\frac{e^2 q^2}{4m} \frac{\tau\omega^2h_xh_y}{(\omega^2-\omega_q^2)^2+ (\omega/\tau)^2}\Re\left[E_{0x}E_{0y}^*
  \right](M_x\hat{\vb*{e}}_x)\\
  &\ \ \ \ \ \ +\frac{e^2 q^2}{4m}\Re\left[
  \frac{i\omega\tau}{(i\omega-1/\tau)(\omega^2-\omega_q^2 -i\omega/\tau)}
  ((h_x\tilde{E}_{0x})(h_y\tilde{E}_{0y}^*)M_y
  +|h_y\tilde{E}_{0y}|^2M_x)
  \right]\hat{\vb*{e}}_y,\\
  \vb*{j}_{DC}^{Lorentz} &=\frac{e^2 q^2}{4m}
  \Re\left[
  \frac{\tau}{\omega^2-\omega_q^2 -i\omega/\tau} |h_y\tilde{E}_{0y}|^2
  \right] M_x\hat{\vb*{e}}_y -\frac{e^2 q^2}{4m}
  \Re\left[
  \frac{\tau}{\omega^2-i\omega/\tau} |h_y\tilde{E}_{0y}|^2
  \right] M_y\hat{\vb*{e}}_x\\
  \vb*{j}_{DC}^{Y} &=\frac{me^3q^4Y}{4\hbar C}\Re\left[
  \frac{i/\omega}{\omega^2-\omega_q^2 +i\omega/\tau} M_x|h_y\tilde{E}_{0y}|^2\hat{\vb{e}}_x  \right].\\
\end{aligned}
\end{equation}
Here $\vb*{j}_{DC}^{density}$ is the photocurrent related with the first order spatial modulation of particle density $\delta n_1$, $\vb*{j}_{DC}^{inertia}$ is the one arising from the 
inertia term $(\uu\cdot \grad)\uu$, $\vb*{j}_{DC}^{Lorentz}$ is the one arising from the Lorentz force term, $\vb*{j}_{DC}^{Y}$ is the one related with the geometrical scalar coefficient $Y$.

By introducing the notation
\begin{equation}
    \tilde{E}_{0i}\tilde{E}_{0j}^*=  \Re\left[\tilde{E}_{0i}\tilde{E}_{0j}^*\right]
    + i\Im\left[\tilde{E}_{0i}\tilde{E}_{0j}^*\right]
    = \mathcal{L}_{ij} - i\epsilon_{ijk}\mathcal{F}_k,
\end{equation}
we can describe the polarization dependence of each term more clearly. Here we have defined $\mathcal{L}_{ij}=\Re\left[\tilde{E}_{0i}\tilde{E}_{0j}^*\right]$ and $\vb*{\mathcal{F}}=\frac{i}{2} \tilde{\E}\times \tilde{\E}^*$, each of which represents the contribution of linearly-polarized light and circularly-polarized light respectively~\cite{Watanabe2021}. 
In particular, the trace of $\mathcal{L}_{ij}$, denoted as $\mathcal{I}=\frac12 \sum_i\mathcal{L}_{ii}$, is proportional to the intensity of an incident light. 
For example, when the incident light is a linearly-polarized light, such as $\tilde{\E}_0=\tilde{E}_0(1,\pm1,0)$, these indicators satisfy $\mathcal{I}=|\tilde{E}_0|^2$, $\mathcal{L}_{xy}=\pm|\tilde{E}_0|^2$ and $\vb*{\mathcal{F}}=0$.
On the other hand, when the incident light is a circularly-polarized light, i.e., $\tilde{\E}_0=\tilde{E}_0(1,\pm i,0)$, these indicators satisfy $\mathcal{I}=|\tilde{E}_0|^2$, $\mathcal{L}_{xy}=0$ and $\vb*{\mathcal{F}}=\pm|\tilde{E}_0|^2\hat{\vb*{e}}_z$.
Using these notations, we can rewrite the above results as

\begin{equation}
\label{eq:photocurrent_contributions_2}
\begin{aligned}
  j_{DC,i}^{BCD} &=\frac{e^3}{2\hbar} \left[ 
  \frac{\tau }{1+(\omega\tau)^2}\sum_{j}
  \left(
  1 + \frac{\omega^2(1+(\omega\tau)^2)h_xh_j}{2\tau^2[(\omega^2-\omega_q^2)^2+(\omega/\tau)^2]}
  \right)
  \epsilon_{ij} D_x L_{xj}
  +\frac{\tau }{1+(\omega\tau)^2}\sum_{j}(1+h_yh_j)\epsilon_{ij} D_y L_{yj}
  \right. \\
  &\quad+\left. \frac{\omega\tau^2}{1+(\omega\tau)^2}
  \left(
  1 - \frac{(\omega^2-\omega_q^2)(1+(\omega\tau)^2) h_xh_y}{2\tau^2[(\omega^2-\omega_q^2)^2+(\omega/\tau)^2]}
  \right)
  \mathcal{F}_z D_x\delta_{ix}
  +\frac{\omega\tau^2}{1+(\omega\tau)^2}(1+h_xh_y)\mathcal{F}_z D_y\delta_{iy}
  \right]\\
  &\quad + \frac{e^5n_0q^2}{4\hbar mC} \frac{1}{(1+(\omega\tau)^2)((\omega^2-\omega_q^2)^2+ (\omega/\tau)^2)}
  \Biggl[
  \tau\omega_q^2 h_xh_y\mathcal{L}_{xy}
  - \omega(1 +\tau^2(\omega^2-\omega^2_q))h_xh_y\mathcal{F}_z
   \\
  &\quad
  +\tau\omega_q^2\left( \frac{mq}{en_0} \right)^2 M_yh_y^2\mathcal{L}_{yy}
  \Biggl]D_y\delta{iy},\\
  \vb*{j}_{DC}^{density} &=\frac{e^2q^2}{4m}
  \left[\frac{\tau(\omega^2-\omega_q^2)h_xh_y}{(\omega^2-\omega_q^2)^2+(\omega/\tau)^2} \mathcal{L}_{xy}
    +\frac{\omega h_xh_y}{(\omega^2-\omega_q^2)^2+(\omega/\tau)^2}\mathcal{F}_z
	\right]\vb*{M}+\frac{e^2q^2}{2m} \frac{\omega h_xh_y}{(\omega^2-\omega_q^2)^2+ (\omega/\tau)^2} 
	\mathcal{F}_zM_x\hat{\vb{e}}_x,\\
  &\quad
  +\frac{e^2q^2}{4m}\frac{1}{(1+(\omega\tau)^2)((\omega^2-\omega_q^2)^2+ (\omega/\tau)^2)}
  \left[
  \omega(1+\tau^2(\omega^2-\omega_q^2))(h_y^2\mathcal{L}_{yy}M_x-h_xh_y\mathcal{L}_{xy}M_y)
  -\tau\omega_q^2 h_xh_y\mathcal{F}_z M_x
  \right]\hat{\vb{e}}_y,\\
  \vb*{j}_{DC}^{inertia} &=\frac{e^2 q^2}{4m} \frac{\omega^2\tau h_xh_y}{(\omega^2-\omega_q^2)^2+ (\omega/\tau)^2}
  \mathcal{L}_{xy}M_x\hat{\vb{e}}_x\\
  &\quad
  +\frac{e^2q^2}{4m}\frac{\omega\tau}{(1+(\omega\tau)^2)((\omega^2-\omega_q^2)^2+ (\omega/\tau)^2)}\left[
  \omega(1+\tau^2(\omega^2-\omega_q^2))(h_y^2\mathcal{L}_{yy}M_x+h_xh_y\mathcal{L}_{xy}M_y)
  +\tau\omega_q^2 h_xh_y\mathcal{F}_z M_y
  \right]\hat{\vb{e}}_y,\\
  \vb*{j}_{DC}^{Lorentz} &=\frac{e^2 q^2}{4m}\frac{\tau(\omega^2-\omega_q^2)h_y^2}{(\omega^2-\omega_q^2)^2+(\omega/\tau)^2} \mathcal{L}_{yy}M_x \hat{\vb*{e}}_y
  -\frac{e^2 q^2}{4m} \frac{\tau^3h_y^2}{1+(\omega\tau)^2}
  \mathcal{L}_{yy}M_y\hat{\vb*{e}}_x,\\
  \vb*{j}_{DC}^{Y} &=\frac{me^3q^4Y}{4\hbar C} \frac{1/\tau}{(\omega^2-\omega_q^2)^2+ (\omega/\tau)^2}\mathcal{L}_{yy}M_x\hat{\vb{e}}_x.\\
\end{aligned}
\end{equation}

\subsection{III. Comment on the viscous effect}

In the hydrodynamic regime, viscosity is one of key ingredients to characterize electron fluids. Actually, a lot of recent works have ever been devoted to measurements of the signature of viscosity in nonlocal transport phenomena~\cite{Bandurin2016,Moll2016,Gooth2018,Levin2018,Gusev2018,Gusev2018-2,Berdyugin2019}.
The hydrodynamic equations obtained in Ref.~\cite{Toshio2020,Sano2021} do not include explicitly the contribution of the viscosity since they focus on the formulation in the limit of $(\omega\tau_e)^{-1}\to 0$, while viscosity is a correction term in the first order of $\tau_e^{-1}$. Here $\omega$ is the characteristic scale of the frequency we discuss. However, the viscous effects on our results are worth discussing by introducing viscosity terms into the theory phenomenologically. In what follows, we briefly discuss the viscous effects especially at the level of linear responses, which is enough to consider the modification of the QNLH term $\vb*{j}_{DC}^{BCD}$.

First we can incorporate the viscous effects into our theory by adding the terms $\nu \Delta u +\zeta \grad (\div\uu)$
to the right side of Eq.~\eqref{eq:hydro_perturbative}. Here $\nu$ is the kinematic viscosity, and $\zeta$ is the bulk viscosity (devided by the mass density $mn$).
As easily checked, we can modify the results in Eq.~\eqref{eq:first_order} by replacing the dissipation rate $1/\tau$ appearing in the prefactor $(\omega^2-\omega_q^2 -i\omega/\tau)^{-1}$ with $1/\tau+(\nu+\zeta) q^2$ for 
the case of longitudinal waves ($\E_0\parallel \hat{\vb*{e}}_x$). 
Clearly, this means that viscosity of electron fluids gives rise to a spatial dispersive contribution in the lifetime of plasmons. When we assume that $\tau=1\times10^{-12}\ \mathrm{s}^{-1}$ and $\nu=1\times10^{-1} \ \mathrm{m^2s^{-1}}$~\cite{Bandurin2016}, the viscosity effect becomes comparable or superior to that of the momentum relaxing scattering rate $1/\tau$ under the condition that $2\pi/q \lesssim 1\ \mu m$.

In particular, focusing on the enhancement factor $\beta_\omega$ in the main text, electron viscosity modifies the factor as

\begin{equation}
      \beta_\omega=
  1 - \frac{\tilde{\omega}^2(1+\tilde{\tau}^2\tilde{\omega}^2)h_x^2}{2[\tilde{\tau}^2(\tilde{\omega}^2-1)^2+\tilde{\omega}^2]}
  \ \ \longrightarrow \ \ 
  1 - \frac{\tilde{\omega}^2(1+\tilde{\tau}^2\tilde{\omega}^2)(\tilde{\tau}_v/\tilde{\tau})h_x^2}{2[\tilde{\tau}_v^2(\tilde{\omega}^2-1)^2+\tilde{\omega}^2]},
\end{equation}
leading to the broadening of the peak width $1/\tilde{\tau}\ \to \ 1/\tilde{\tau}_v$ under the resonant condition, and the suppression of the peak amplitude $|\beta_{\omega_q}| \simeq |h_x\tilde{\tau}|^2/2\  \to \ h_x^2\tilde{\tau}\tilde{\tau}_v/2
$. Here we have introduced a new dimensionless parameter $\tilde{\tau}_v=\omega_q(1/\tau+(\nu+\zeta)q^2)^{-1}\ (<\tilde{\tau})$.

\end{document}